\documentclass[12pt]{article}
\usepackage[dvips]{epsfig}

\newcommand{\be}{\begin{equation}}
\newcommand{\ee}{\end{equation}}
\newcommand{\bn}{\begin{eqnarray}}
\newcommand{\en}{\end{eqnarray}}

\def \ps
{p\!\!\!/}
\def \pa
{A\!\!\!/}
\def \pb
{B\!\!\!/}

\newcommand{\beq}{\begin{equation}}
\newcommand{\eeq}{\end{equation}}

\begin{document}

\title{\textbf{Effective Action for QED with Fermion Self-Interaction in D=2 and
D=3 Dimensions}}
\author{E. M. C. Abreu, D. Dalmazi, A. de Souza Dutra and Marcelo Hott \\
\textit{UNESP - Campus de Guaratinguet\'a - DFQ} \\
\textit{Av. Dr. Ariberto Pereira da Cunha, 333} \\
\textit{CEP 12516-410 - Guaratinguet\'a - SP - Brazil.} \\
\textsf{E-mail: everton@feg.unesp.br, dalmazi@feg.unesp.br,
dutra@feg.unesp.br }\\
\textsf{and hott@feg.unesp.br}}
\date{\today}
\maketitle

\begin{abstract}
In this work we discuss the effect of the quartic fermion self-interaction of Thirring
type in QED in $\,D=2$ and $\,D=3$ dimensions. This is done through the computation of
the effective action up to quadratic terms in the photon field. We
analyze the corresponding nonlocal photon propagators nonperturbatively in $%
\frac{k}{m}$, where $k$ is the photon momentum and $m $ the fermion mass. The poles of
the propagators were determined numerically by using the Mathematica software. In
$D=2$ there is always a massless pole whereas for strong enough Thirring coupling a
massive pole may appear . For $D=3$ there are three regions in parameters space. We
may have one or two massive poles or even no pole at all. The inter-quark static
potential is computed analytically in $D=2$. We notice that the Thirring interaction
contributes with a screening term to the confining linear potential of massive
QED$_{2}$. In $D=3$ the static potential must be calculated numerically. The screening
nature of the massive QED$_{3}$
prevails at any distance, indicating that this is a universal feature of $%
D=3 $ electromagnetic interaction. Our results become exact for an infinite
number of fermion flavors. \newline
\textit{PACS-No.:} 11.15.Bt , 11.15.-q
\end{abstract}



\newpage

\section{Introduction}


In two dimensions, bosonization is a powerful technique used in a variety of examples
\cite{livroElcio,Jackiw,Coleman}. In the last years there has been many attempts to
generalize those ideas to higher dimensions [4-10]. For instance, one can derive an
effective action by integrating out the fermion degrees of freedom and studying the
physical properties of the resulting bosonic effective theory. Such approach has been
used in [4,13,28] to show that the static potential in QED$_3$ is of screening type.
In [4] we have used the perturbative path integral bosonization in both $D=2$ and
$D=3$ QED. It is remarkable that in $QED_{2}$ at the quadratic approximation in the
gauge fields but without any expansion in $k/m$, there is only massless poles
\cite{nossoartigo}, which is in agreement with what has been observed
in \cite{gross}, but differs from the result obtained through perturbative ($%
\,m/e\,$) calculation of \cite{adam}. In three dimensions it was shown that
there is a massive excitation which depends on the dimensionless parameter $%
16\pi \,m/e^{2}$ and a simple approximated expression for this function has
been found \cite{nossoartigo}. This in fact generalizes the calculations of
\cite{schaposnik}, which were obtained at leading order of the derivative
expansion, and that of \cite{cesar99} carried out at a higher order in $k/m$%
, which in its turn is related to consistent higher derivative actions \cite
{cesar97, jackiw99}.

The aim of this work is to analyze the influence of adding a Thirring term
to QED in the static potential as well as in the particle content of the
theory. In particular, we conclude that such term does not change the large
distance physics.

We start by introducing the notation which will be used in both $D=2$ and $%
D=3$. The generating functional for QED with Thirring self-interaction is
given by
\begin{eqnarray}
Z &=&\int \mathcal{D}A_{\mu }\,\mathcal{D}\psi \,\mathcal{D}\bar{\psi}%
\,exp\left\{ i\,\int \,d^{D}x\,\,\left[ -\frac{1}{4}\,F_{\mu \nu }^{2}\,+\,%
\bar{\psi}^{j}\,(i\,\partial \!\!\!/\,-\,m\,-\, \frac{e}{\sqrt{N}}\,\pa)\psi
^{j}\right. \right.  \nonumber \\
&&\left. \left. -\,\frac{g^{2}}{2N}\,(\psi ^{j}\gamma ^{\mu }\psi
^{j})^{2}\,\right] \,\right\} \ .
\end{eqnarray}

\noindent where $N$ is the number of fermion flavors. It is convenient to
introduce an auxiliary vector field $B_{\mu }$ and work with the physically
equivalent generating functional:

\begin{eqnarray}
Z &=&\int \mathcal{D}A_{\mu }\,\mathcal{D}B_{\mu }\mathcal{D}\psi \,\mathcal{%
D}\bar{\psi}\,exp\left\{ \,i\,\int \,d^{D}x\,\,\left[ -\frac{1}{4}\,F_{\mu
\nu }^{2}\,+\,{\frac{1}{2}}B_{\mu }B^{\mu }+\right. \,\right.  \nonumber \\
&&\left. \left. +\bar{\psi}^{j}\,(i\,\partial \!\!\!/\,-\,m\,-\,\frac{e}{%
\sqrt{N}}\, \pa-\, \frac{g}{\sqrt{N}}\,\pb)\psi ^{j}\right] \right\} \ ,
\end{eqnarray}

After integration over the fermionic fields we obtain

\begin{eqnarray}
Z &=&\int \mathcal{D}A_{\mu }\mathcal{D}B_{\mu }exp\left\{ \,i\,\int
\,d^{n}x\,\left[ \,-\frac{1}{4}\,F_{\mu \nu }^{2}\,+\,{\frac{1}{2}}B_{\mu
}B^{\mu }\right] \right\} \times  \nonumber \\
&&\times det\left[ i\partial \!\!\!/\,-\,m\,-\,{\frac{1}{\sqrt{N}}}(e\,\pa%
\,+\,g\,\pb)\,\right] ^{N}
\end{eqnarray}
The fermion determinant can be evaluated perturbatively in $1/N$ and Furry's
theorem guarantees that only even number of vertices contribute. Since each
vertex is of order $1/\sqrt{N} $ the leading contribution with two vertices
will be $N$-independent. The next to leading contribution with four vertices
is of order $1/N$ and will be neglected henceforth. Therefore, at leading
order in $1/N$, we have the quadratic effective action:
\begin{eqnarray}
S_{eff}^{(2)}[B,A] &=&\frac{1}{2}\int \;\frac{d^{D}k}{(2\pi )^{D}}\;\left\{ -%
{\tilde{A}}_{\mu }(-k)\left[ g^{\mu \nu }k^{2}-k^{\mu }k^{\nu }\right] {%
\tilde{A}}_{\nu }(k)\,+\,{\tilde{B}}^{\mu }(k){\tilde{B}}_{\mu }(-k)\right.
\nonumber \\
&&\left. \,+\,(e\,\tilde{A}_{\mu }\,+\,g\,\tilde{B}_{\mu })(k)\Pi ^{\mu \nu
}(e\,\tilde{A}_{\nu }\,+\,g\,\tilde{B}_{\nu })(-k)\right\} \;\;,
\end{eqnarray}

\noindent where ${\tilde{A}}_{\mu}(k)$ and ${\tilde{B}}_{\mu}(k)$ represent
the Fourier transformations of $A_{\mu}(x)$ and ${B}_{\mu}(k)$ respectively,
and $\Pi^{\mu\nu}$ is the polarization tensor:

\begin{equation}
\Pi ^{\mu \nu }(k)=i\int \;\frac{d^{D}p}{(2\pi )^{D}}\;tr\left[ \frac{1}{\ps %
-m+i\epsilon }\gamma ^{\mu }\frac{1}{(p\!\!\!/+k\!\!\!/)-m+i\epsilon }\gamma
^{\nu }\right]
\end{equation}
the action $S_{eff}^{(2)}$ $[B,A]$ is exact in the $N\rightarrow \infty $
limit.

In order to proceed further we have to calculate $\Pi^{\mu\nu}$ which
depends on the dimensionality of the space-time.

\section{Effective potential in D=2}

Using dimensional regularization we have
\begin{equation}
\Pi _{D=2}^{\mu \nu }(k)=\,g^{2}\,\tilde{\Pi}(k^2)\,\theta ^{\mu \nu },
\end{equation}
where $\theta ^{\mu \nu }=g^{\mu \nu }-\frac{k^{\mu }k^{\nu }}{k^{2}}$ and

\begin{equation}
{\tilde{\Pi}}(k^{2})={\frac{1}{\pi }}\left[ 1+\frac{1}{2}\frac{4{m^{2}}/k^{2}%
}{(1-{4{m^{2}}/k^{2}})^{1/2}}\ln {\frac{(1-{4{m^{2}}/k^{2}})^{1/2}+1}{(1-{4{%
m^{2}}/k^{2}})^{1/2}-1}}\right] \;\;.  \label{7}
\end{equation}
Plugging back in $S_{eff}^{(2)}[B,A]$ and performing the Gaussian integral
over $B^{\mu }$ we end up with the gauge invariant effective action for the
gauge field:
\begin{equation}
S_{eff}^{(2)}[A]=\frac{1}{2}\int \;\frac{d^{2}k}{(2\pi )^{2}}\;{\tilde{A}}%
_{\mu }(-k)\left[ -k^{2}+e^{2}\frac{\tilde{\Pi}}{1\,+\,g^{2}\tilde{\Pi}}%
\right] \theta ^{\mu \nu }\,{\tilde{A}}_{\nu }(k)\;\;.  \label{4}
\end{equation}

If $g\rightarrow 0$ we reproduce the $QED_{2}$ result of \cite{nossoartigo},
and when $m\,\rightarrow \,0$ we recover the result of the
Schwinger-Thirring model \cite{belve1}. Introducing a gauge fixing term we
can obtain the photon propagator whose gauge invariant piece is given by:

\begin{equation}
{D^{\parallel }}_{\mu \nu }(k)=\frac{(\,1\,+\,\tilde{\Pi}\,g^{2}\,)}{%
4\,m^{2}\left[ g^{2}\,\tilde{\Pi}(z-a)\,+\,z\right] }\,g_{\mu \nu },
\label{prop2d}
\end{equation}
where we define the dimensionless quantities which will be used also in $D=3$%
:

\begin{eqnarray}  \label{11}
z&=&\frac{k^2}{4\,m^2} \;\;,  \nonumber \\
a&=&\frac{e^2}{4\,m^2\,g^2}\;\;.
\end{eqnarray}

\noindent Notice that $g$ is another dimensionless quantity in $D=2$.

As already stressed in \cite{nossoartigo} the expression (\ref{7}) for $%
\tilde{\Pi}$ is correct for $k^{2}<0$ and we have used it to check that $%
D_{\mu \nu }$ is causal and no tachyonic poles appear. In analyzing the
particle content of $S_{eff}^{(2)}[A]$ we restrict ourselves to the region $%
0\leq z<1$, which is below the pair creation threshold $(z=1)$. In that
region $\tilde{\Pi}$ must be continued to

\begin{equation}
\tilde{\Pi}(z)\,=\,\frac{1}{\pi}\,\left[ \;1\,-\,\frac{1}{[z(1-z)]^{1/2}}%
\,\arctan\, \sqrt{\frac{z}{1-z}}\; \right] \qquad ; \qquad 0 \leq z < 1\;\;.
\end{equation}

For $z\rightarrow 0$ , $\tilde{\Pi}$ becomes linear in $z$ and therefore,
for arbitrary values of the dimensionless parameters $a$ and $g$, we always
have a massless simple pole $k^{2}=0$ as in pure QED$_2$ [4]. Since $\tilde{%
\Pi}\leq 0$ for $0\leq z<1$ it is clear that $z\,+\,g^{2}\,\tilde{\Pi}%
\,(z-a) $ will never vanish for $a\geq 1$ and we are left with only a
massless pole . For $a<1$ there is always a massive pole at $\frac{k^{2}}{%
2m^{2}}\,=\,M_{2D}(a,g)$, which was numerically evaluated and plotted in Figure 1 in
the region: $0.1\leq a\leq 0.93$ and $0.25<g<1.5$.

As we see in Figure 1, when we decrease the Thirring coupling the mass of the pole
tends to reach the pair creation threshold $z\to 1^{-} $ and becomes nonphysical.

At this point, we observe that the effect of the Thirring self-interaction
in the pole structure of $QED_{2}$ is to introduce a massive pole if the
Thirring coupling is strong enough, i.e., $mg^{2}/4\pi >e^{2}/16\pi m$. The
massless pole of pure $QED_{2}$ with massive fermions remains untouched for
any value of the coupling $e$ and $g$, wich is compatible with a recent
study \cite{belve2}.

Now two comments are in order. First, though the Thirring interaction may
introduce a mass in the photon propagator the gauge symmetry is not broken.
As we see in (8) this is only possible because the action is nonlocal. If we
try to make it local, for instance by making a derivative expansion for $%
m\to \infty $. We will miss the massive pole since the Thirring contribution
will be neglected at first order in the derivative expansion. The massive
pole will be only seen in the next to leading order in a higher derivative
theory. Second, for any $g\ne 0 $ it is always possible to find a value for $%
k^2$ such that $\,1\,+\,\tilde{\Pi}\,g^{2}\, = \, 0$ and therefore $%
D^{\parallel }_{\mu \nu }\, =\, 0$. That is, the Thirring self-interaction
originates a region in momentum space (hyperboloid) of forbidden momenta. We
are not aware of similar observations in the literature and we do not have a
deeper understanding of this fact.

Next we analyze the effect of Thirring interaction for the effective
potential between two static charges. We assume that two charges $Q$ and $-Q$
are located at $x=L/2$ and $x=-L/2$ From the equation of motion coming from $%
S_{eff}^{(2)}[A]$ we obtain the potential produced by the positive charge:

\begin{equation}
A_{\mu}(x) = \int \; \frac{d^{2}k}{(2 \pi)^2}\int d^{2}x^{\prime}
D^{\|}_{\mu \nu}(k) e^{i k (x - x^{\prime})} J^{\nu}(x^{\prime}) ,
\end{equation}

\noindent where

\begin{equation}
J^{\nu} (x^{\prime}) = Q \delta({x_{1}}^{\prime} - {\frac{L }{2}})
\delta^{\nu 0}
\end{equation}
\label{current}

\noindent and $D_{\mu \nu }^{\Vert }$ is given in (\ref{prop2d}). The only
non-vanishing component of the potential is $A_{0}$: which can be obtained
analytically through a contour integral:

\begin{equation}
V(L)\,=\,A_{0}\,(x=-L/2)\,=\,9\,\left[ \frac{L}{2(1+\frac{2ag^{2}}{3\pi })}%
\,+\,\frac{2a}{m\sqrt{u}}\,\frac{e^{-2m\sqrt{u}L}}{(u-a)^{2}\tilde{\Pi}(u)-a}%
\right]
\end{equation}
where $u$ is the non-vanishing solution of
\begin{equation}
g^{2}{\tilde{\Pi}}(z)(z-a)+z\,=\,0,  \label{v2}
\end{equation}
with ${\tilde{\Pi}}$ given in (12). The \textit{screening} effect, second
term in $V(L)$, only exists for $a<1$ and is a pure consequence of the
Thirring coupling. At large distances the \textit{confining} nature massive $%
QED_{2}$ prevails and the influence of the Thirring term fades away.

\section{The massive poles in D=3}

Using a parity and gauge invariant regulator, we obtain in this case:
\[
\Pi ^{\mu \nu }(k)= \frac{f_{1}(k^2)}{8\pi }\,i\,\epsilon ^{\mu \nu \rho
}k_{\rho }+\frac{f_{2}(k^2)}{16\pi \,m}\,k^{2}\,\theta ^{\mu \nu }\,,
\]

\noindent where, for $0\leq \,z\,<\,1$, we have the parametric functions:
\begin{equation}
f_{1}\left( z\right) \,=\,-\frac{1}{z^{1/2}}\ln \left( \frac{1+z^{1/2}}{%
1-z^{1/2}}\right) ;\,\,f_{2}\left( z\right) \,=\,\frac{1}{z}\left[ 1+\left(
\frac{1+z}{2}\right) f_{1}\right] .
\end{equation}

\noindent Again, plugging back in $S_{eff}^{\left( 2\right) }[B,A]$ and
integrating over $B_{\mu }$ we obtain :
\begin{eqnarray}
S_{eff}^{(2)}[A] &=&\int \;\frac{d^{3}k}{(2\pi )^{3}}\;{\tilde{A}}_{\mu }(-k)\left\{
\left[ -k^{2}+e^{2}\left(
1+\frac{k^{2}f_{2}-k^{2}g^{2}f_{1}^{2}}{(g^{2}k^{2}f_{2}+1)^{2}-g^{4}k^{2}{f
_{1}}^{2}}\right)
\right] \theta ^{\mu \nu }\right.  \nonumber \\
&&\left. +i\,\frac{e^{2}\,f_{1}}{(g^{2}k^{2}{f_{2}}+1)^{2}-g^{4}k^{2}{%
f_{1}}^{2}}\epsilon ^{\mu \nu \rho }k_{\rho }\right\} {\tilde{A}}_{\nu }(k),
\end{eqnarray}

\noindent Once again we recover the pure $QED_{3}$ result for $g\rightarrow \,0$. As
in $D=2$ case $S_{eff}^{\left(2\right)}[A]$ is gauge invariant and, after introducing
a gauge fixing term, we can write the gauge invariant piece of the ''photon´´
propagator as
\begin{equation}
D_{\mu \nu }\,=\,\frac{\mathcal{N}\left( a,b,z\right) }{4\,m^{2}D_{+}\,D_{-}}%
\,g_{\mu \nu }\,-\,\frac{i\,e^{2}f_{1}}{32\,\pi \,m\,k^{2}D_{+}\,D_{-}}%
\,\epsilon _{\mu \nu \rho }k^{\rho }.  \label{C}
\end{equation}

\noindent where $a$ is defined as in (\ref{11}) and
\begin{eqnarray}  \label{19}
b\,&=&\,\frac{m\,g^{2}}{4\pi }\;\;;  \nonumber \\
\,\mathcal{N\,}\,&=&\,\left( 1+b\,z\, 16 \pi m f_2\right) \left[ b\,\, 16
\pi m f_2\left( a-z\right) \,-\,1\right] \,-\,b^{2}\,8 \pi
f_1\left(a-z\right) ;
\end{eqnarray}

\begin{eqnarray}  \label{20}
D_{\pm }\,&=&\,b\left( \,a-z\right) \,G_{\pm }\,-\,\sqrt{z}\,\;\;; \\
\,G_{\pm }\,&=&\,\frac{1}{\sqrt{z}}\left[ 1+\frac{\left( 1\,\pm \,\sqrt{z}%
\right) ^{2}}{2}\,f_{1}\right] .
\end{eqnarray}

\noindent Both $G_{\pm }$ are monotonically decreasing functions which
satisfy:
\begin{equation}  \label{21}
1\leq \,G_{-}\,\leq \,2 \qquad \mbox{and} \qquad G_{+}\,\leq \,-2\;\;.
\end{equation}
for $z \leq 1$. Therefore when looking for the poles of the propagator:
\begin{equation}
D_{+}\,D_{-}\,=\,0,
\end{equation}
\noindent it is natural to split the analysis in two regions:

\subsubsection{$\,a<1,\,$ $\left( \frac{e^{2}}{16\pi m}\,<\,\frac{m\,g^{2}}{%
4\pi }\right) $}

In this case we can have $D_{+}\,=\,0$ for some $z>a$ and $D_{-}\,=\,0$ for
some $z\,<\,a$. Indeed we have always been able to find massive poles in
both regions ($z>a$ and $z<a$) for arbitrary values of $b$ and $0<z<1 $.

We might say that these poles have distinct origins. The first one is due to the
fermion self-interaction, whereas the second one has its origin due to the dynamically
generated Chern-Simons term. This can be seen if one works with the reducible
representation for the fermion field. In this case the parity-odd term is not
generated and, as a consequence, the gauge invariant piece of the propagator is given
by

\[
D_{\mu \nu }\,=\,\frac{(1+b\,z\,16\pi m {f}_{2})}{4\,m^{2}\,z\,[b\,\,16\pi m
{f}_{2}(a-z)-1]}\,g_{\mu \nu }\,\,\,\,\,.
\]

Since $\, {f}_{2}\leq 0$ for $0\leq z<1$ one can check that there is always
a massless pole and another one at $z_{o}>a$ which satisfies

\[
b\,\,16\pi m {f}_{2}(z_{o})\,(a-z_{o})-1=0\,\,\,\,.
\]

\noindent In pure $QED_{3}$ with reducible representation no mass is
dynamically generated for the gauge field, consequently the massive pole we
have found above has its origin in the Thirring term.

\subsubsection{$a\thinspace \geq \,1,\,\,\left( \frac{e^{2}}{16\pi m}
\,\,\geq \,\frac{m\,g^{2}}{4\pi }\right) $}

In this $QED_{3}$ dominated region we can only have poles from $D_{-}\,=\,0$
or equivalently:
\begin{equation}  \label{B}
b\,G_{-}\,=\,\frac{\sqrt{z}}{a-z}.
\end{equation}

\noindent Since the right hand side of (\ref{B}) is a monotonically
increasing function of $z$ in the range $0 < z < 1 $ and $G_{-}$ is limited
according to (\ref{21}), it is clear that there are no solutions for $%
D_{-}\,=\,0$ whenever $b\geq \frac{1}{a-1}$; \textit{i.\thinspace e.}, $%
a\geq \,1+\frac{1}{b}$. Therefore in terms of $QED_{3}$ and Thirring
couplings, if
\begin{equation}
\frac{e^{2}}{16\,\pi \,m}\,\geq \,\frac{m\,g^{2}}{4\,\pi }\,+\,1,
\end{equation}

\noindent the propagator (\ref{C}) has no poles whatsoever, this may happen
as an artifact of the approximation. It is possible that, as at this order
no poles do appear, the usually negligible next perturbative contribution
could introduce back the massive pole. We have found numerically that the
above bound indeed exists. For any $b\,<\,\frac{1}{a-1}$ we have always been
able to find one massive pole at $D_{-}\,=\,0$ for arbitrary values of the
parameter $a$ in the region $a>1$.

Summarizing,

\noindent i) If $\frac{e^{2}}{16\,\pi \,m}\,<\,\frac{m\,g^{2}}{4\,\pi }\,$,
then two massive poles ($D_{\pm }\,=\,0$) are present

\noindent ii) If $\frac{m\,g^{2}}{4\,\pi }\leq \,\frac{e^{2}}{16\,\pi \,m}<\,%
\frac{m\,g^{2}}{4\,\pi }\,+\,1$, just one pole ($D_{-}\,=\,0$) appears.

\noindent iii) Finally, if $\frac{e^{2}}{16\,\pi \,m}\geq \,\frac{m\,g^{2}}{%
4\,\pi }\,+\,1$, no poles appear at all.

Now two remarks follow. First, concerning the dependence on the $QED_{3}$
and Thirring couplings on the massive pole found from $D_{-}\,=\,0$, for $%
b<\,\frac{1}{a-1}$, we have found numerically and it is plotted in Figure 2, that the
mass increases along with the $QED_{3}$ coupling and decreases for
growing Thirring coupling. If we take both small $m g^{2}$ and large $\frac{%
e^{2}}{4 m}$ we tend to violate the condition $\frac{e^{2}}{16\,\pi \,m}<\,%
\frac{m\,g^{2}}{4\,\pi }\,+\,1$ and the pole tends to go beyond the pair
creation threshold ($z \geq 1$) as we see on the top of the hill in Figure
2. The second comment regards the pure $QED_{3}$ limit\thinspace \thinspace (%
$g\,\rightarrow \,0$) for which there is still a region without poles in the
propagator, \textit{i. e.}, $\frac{e^{2}}{16\,\pi \,m}\,\geq \,\,1$. This
seems to have gone unnoticed in the literature \cite{elcio2}\cite{sghosh}%
\cite{nossoartigo}. Sometimes the quadratic approximation for $%
S_{eff}^{(2)}[A]$ is called a small coupling approximation in the
literature, thus one might argue that our calculations only make sense for
small coupling $e$, such that we are below the bound $e^{2\,}\,=\,16\,\pi
\,m $. This is certainly sensible at the leading order in the derivative
expansion, as in \cite{elcio} since $m\,\rightarrow \,\infty $ but it is not
true in general. In particular, for $QED_3$ with large number of flavors, we
have argued that the quadratic approximation for the effective action
corresponds to the leading $1/N$ contribution and no restriction is required
on $e$ or $m $, therefore the problem persists.

Similarly to the $D=2$ case we now move to the calculation of the effect of
the Thirring self-interaction on the potential between two static charges $%
+Q $ and $-Q$ located at $\left( x,y\right) _{+}=\left( \frac{L}{2}%
,\,0\right) $ and $\left( x,y\right) _{-}=\left( -\,\frac{L}{2},\,0\right) $%
. That is, the current of the positive charge is
\begin{equation}
J^{\nu}\left( x^{\prime}\right) \,=\,Q\,\delta \left( x^{\prime}\,-\,\frac{L%
}{2}\right) \delta \left( y^{\prime}\right) \,\delta^{\nu 0},
\end{equation}

\noindent the potential produced by the above charge is:
\begin{equation}
A_{0}(x) \,=\,\int d^{3}x^{\prime}\,\int\,\frac{d^{3}k}{( 2\pi)^{3}}%
\,e^{i\,k(x-x^{\prime})}D_{0\alpha}(k) J^{\alpha }( x^{\prime})
\end{equation}

\noindent and
\begin{equation}  \label{27}
V\left( L\right) \,=\,A_{0}\left( x=-\frac{L}{2},\,y=0\right) =\,\frac{Q}{%
4\,m^{2}}\int_{0}^{\,\infty }dk\,k\,\frac{J_{0}\left( k\,L\right) \,\mathcal{%
N}}{D_{+}\,D_{-}},
\end{equation}

\noindent where $k=\sqrt{k_{x}^{2}\,+\,k_{y}^{2}}$ and we have used $%
\epsilon_{0\alpha \gamma }\partial ^{\gamma }J^{\alpha }\,=\,0$. The fact
that the current is static gives rise to a $\delta \left( k_{0}\right) $
upon integration over $dx_{0}^{\prime}$. Since $k_{0}=0$ we have $%
k^{2}=-\left( k_{x}^{2}\,+\,k_{y}^{2}\right) $ and $\frac{\,\mathcal{N}}{%
D_{+}\,D_{-}}$ in (\ref{27}) are the continuations of the expressions (\ref
{19}) and (\ref{20}) for the region $k^{2}\,<\,0$ according with the formula
$\ln \left[ \left( 1-\sqrt{z}\right) /\left( 1+\sqrt{z}\right) \right]
=2\,i\,\arctan \left( z\right) $. The Bessel function $J_{0}\left(
k\,L\right)$ appears after integration over the angular variable. Different
from $D=2$ we are no longer able to calculate $V\left( L\right) $
analytically and we have to appeal to a numeric computation as in \cite
{elcio2}. We have plotted the result in Figure 3 for specific values of $%
m,\,a$ and $b$. We have noticed that the screening form of the potential is
insensitive to the parameters $m,\,a$ and $b$, which is quite surprising in
view of our previous analysis of the pole content of the propagator. The
presence of the Thirring self-interaction seems to be irrelevant for the
static potential even at small distances. Our conclusion is in disagreement
with \cite{sghosh} (see also \cite{gaete}) who claims that because of the
Thirring term, a repulsive barrier appears at low distances. The author of
\cite{sghosh} makes use of the derivative expansion in the quadratic action $%
S_{eff}^{(2)}[A]$, which is presumably a good approximation for large
fermion masses $\frac{k}{2\,m}\rightarrow 0$. We have also checked that $%
V\left( L\right) $ keeps its screening shape, even for large masses, for any distance
$\, L \, $. Changing the values of the couplings $g $ and $e$ will not change the
shape of $V(L)$ either (see Figure 3 for the typical shape). The point is that the
rapid oscillations of the Bessel function washes out any
detail of the photon propagator leading always to a screening potential in $%
D=3$. Finally, similarly to $D=2$ it is always possible to find $k^2$ such
that $\, \mathcal{N}(a,b,z) =0 \, $ and the symmetric part of the photon
propagator ( see (18) ) will vanish for those special values of momenta.

\section{Acknowledgments}

E. M. C. A. is financially supported by Funda\c{c}\~{a}o de Amparo \`{a}
Pesquisa do Estado de S\~{a}o Paulo \textbf{(FAPESP)} (grant 99/03404-6).
This work was partially supported by \textbf{CNPq} and \textbf{FAPESP},
brazilian research agencies.\pagebreak

\newpage

\begin{figure}[tbp]
\begin{center}
\begin{minipage}{20\linewidth}
\epsfig{file=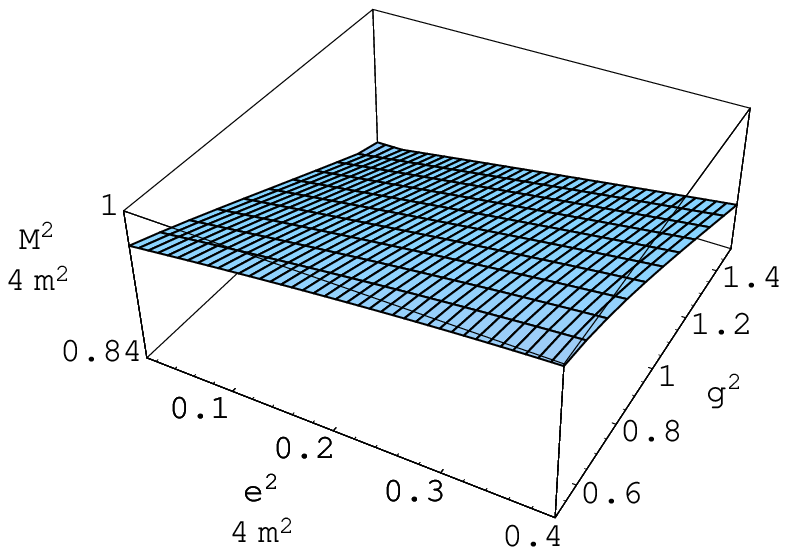}
\end{minipage}
\end{center}
\caption{Photon effective mass in $D=2$ for $a<1$ and $0.25<g<1.5$. } \label{fig:fig1}
\end{figure}

\begin{figure}[tbp]
\begin{center}
\begin{minipage}{20\linewidth}
\epsfig{file=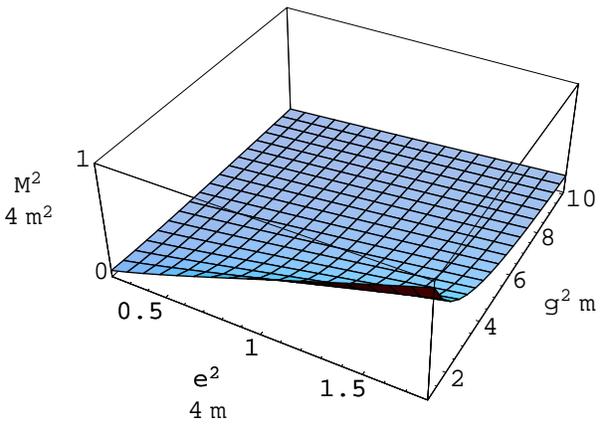}
\end{minipage}
\end{center}
\caption{Behavior of the photon effective mass in $D=3$.} \label{fig:fig2}
\end{figure}

\begin{figure}[tbp]
\begin{center}
\begin{minipage}{20\linewidth}
\epsfig{file=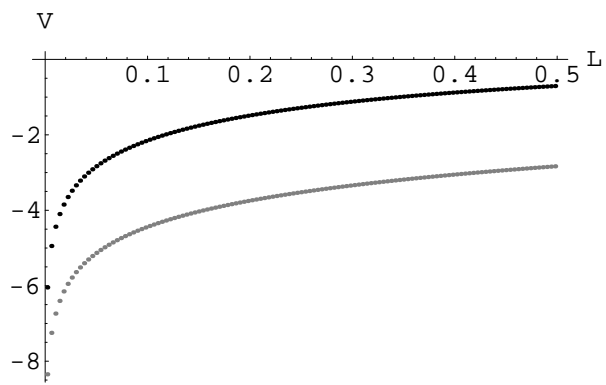}
\end{minipage}
\end{center}
\caption{Inter-fermion effective potential for $a=1/2$, $b=3$, $m=1$ (black dotted
line) and $m=0.1$ (grey dotted line). } \label{fig:fig3}
\end{figure}

\end{document}